\documentclass[aps,twocolumn,superscriptaddress]{revtex4-1}
\usepackage[utf8]{inputenc}
\usepackage[T1]{fontenc}
\usepackage[english]{babel}
\usepackage{graphicx}
\usepackage{mathtools}

\newcommand{\la}{\langle}
\newcommand{\ra}{\rangle}
\newcommand{\be}{\begin{equation}}
\newcommand{\ee}{\end{equation}}

\begin{document}

\title{Experimental validation of  fully quantum fluctuation theorems}

\author{Kaonan Micadei}

\affiliation{Institute for Theoretical Physics I, University of Stuttgart, D-70550 Stuttgart, Germany}

\author{John P. S. Peterson}

\affiliation{Centro Brasileiro de Pesquisas F\'isicas, Rua Dr. Xavier Sigaud 150,
22290-180 Rio de Janeiro, Rio de Janeiro, Brazil}
\affiliation{Institute for Quantum Computing and Department of Physics and Astronomy, University of Waterloo, Waterloo N2L 3G1, Ontario, Canada}

\author{Alexandre M. Souza}

\affiliation{Centro Brasileiro de Pesquisas F\'isicas, Rua Dr. Xavier Sigaud 150,
22290-180 Rio de Janeiro, Rio de Janeiro, Brazil}

\author{Roberto S. Sarthour}

\affiliation{Centro Brasileiro de Pesquisas F\'isicas, Rua Dr. Xavier Sigaud 150,
22290-180 Rio de Janeiro, Rio de Janeiro, Brazil}

\author{Ivan S. Oliveira}

\affiliation{Centro Brasileiro de Pesquisas F\'isicas, Rua Dr. Xavier Sigaud 150,
22290-180 Rio de Janeiro, Rio de Janeiro, Brazil}

\author{Gabriel T. Landi}

\affiliation{Instituto de F\'isica, Universidade de S\~ao Paulo, C.P. 66318, 05315-970
S\~ao Paulo, SP, Brazil}

\author{Roberto M. Serra}

\affiliation{Centro de Ci\^encias Naturais e Humanas, Universidade Federal do ABC,
Avenida dos Estados 5001, 09210-580 Santo Andr\'e, S\~ao Paulo, Brazil}

\affiliation{Department of Physics, University of York, York YO10 5DD, United
Kingdom}

\author{Eric Lutz}

\affiliation{Institute for Theoretical Physics I, University of Stuttgart, D-70550 Stuttgart, Germany}

\begin{abstract}
Fluctuation theorems are fundamental extensions of the second law of thermodynamics for  small systems. Their general validity arbitrarily far from equilibrium makes them invaluable in nonequilibrium physics. So far, experimental studies of quantum fluctuation relations do not account  for quantum correlations and quantum coherence, two essential quantum properties. We here experimentally verify  detailed and integral fully quantum fluctuation theorems for heat exchange using two quantum-correlated thermal spins-1/2 in a nuclear magnetic resonance setup. We confirm, in particular, individual integral fluctuation relations for   quantum correlations and quantum coherence, as well as for the sum of all  quantum contributions. These refined formulations of the second law are important  for the investigation of fully quantum features in nonequilibrium thermodynamics.
\end{abstract}

\maketitle

A defining property of out-of-equilibrium systems is that they dissipate energy, leading to an irreversible increase of their entropy.
The irreversible entropy production is thus a central quantity of nonequilibrium thermodynamics in  the same way that entropy is a central quantity of equilibrium physics \cite{leb08}. In small systems dominated by thermal or quantum fluctuations, the entropy production $\Sigma$ is a stochastic variable \cite{bus05,sek10}. Detailed fluctuation relations quantify the occurrence of negative entropy production events via the general equality $P(\Sigma)/P(-\Sigma) = \exp(\Sigma)$ for the  distribution $P(\Sigma)$ \cite{sei12,jar11,cil13}. Integral fluctuation theorems of the form $\langle \exp(-\Sigma)\rangle =1$ are obtained after integration over $\Sigma$. Both  relations imply the second law of thermodynamics, $\langle \Sigma \rangle \geq 0$, and are therefore regarded as its far-from-equilibrium generalization. They are among  only few exact equalities known to be valid beyond the linear-response regime \cite{sei12,jar11,cil13}.

A standard procedure to investigate quantum fluctuation theorems, both theoretically and experimentally, is the two-projective-measurement approach \cite{esp09,cam11}.  In this framework, the energy change of a quantum system, and accordingly its stochastic entropy production,  are determined    by projectively measuring the energy at the beginning and at the end of a nonequilibrium process \cite{tal07}. 
Equivalent schemes based on  generalized measurements \cite{hub09,ron14} and Ramsey-like interferometry  \cite{maz13,dor13} have additionally been developed. These   methods have been successfully implemented to test  quantum fluctuation relations  for mechanically  driven \cite{bat14,an15,cer17,smi18,zhang18} and thermally driven \cite{pal18,gomez19} systems, using a variety of experimental platforms, such as nuclear magnetic resonance, ion traps, cold atoms,  nitrogen-vacancy centers and superconducting qubits \cite{bat14,an15,cer17,smi18,zhang18,pal18,gomez19}. However,  due to their  inherent projective nature, they fail to capture  quantum correlations and quantum coherence that may be present in initial and final states of the system. Since these are  two central quantum  features \cite{hor09,str17}, such fluctuation theorems may  be viewed as not fully quantum \cite{jev15,alh16,abe18,par17,Santos19,fra19,man18,kwo19}.  

We here report the first experimental study of fully quantum fluctuation relations for heat exchange between  two initially quantum-correlated qubits  prepared in local thermal states at different  temperatures using nuclear magnetic resonance techniques \cite{oli07,van04}.  After initiating thermal coupling between the two qubits, we analyze the statistics of the  exchanged heat by tracking the evolution
of the global two-qubit state  with the help of quantum state tomography \cite{oli07}. We determine the heat distribution, at any time, during a forward nonequilibrium heat exchange process, as well as during its (time) reverse, both with and without initial quantum correlations between the qubits. In the absence of initial correlations, we verify the detailed fluctuation relation for heat obtained by Jarzynski and  W\'ojcik within the two-projective-measurement scheme  \cite{jar04}. In the presence of initial quantum correlations, we  confirm a modified fluctuation theorem  derived using a dynamic Bayesian network approach that fully accounts for quantum correlations and quantum coherence at all times \cite{mic20}. We further demonstrate the validity  of independent integral fluctuation relations for classical correlations (in the form of a stochastic classical mutual information \cite{nie00}), quantum correlations (in the form of a stochastic quantum mutual information \cite{nie00}) and quantum coherence (by means of a stochastic relative entropy of coherence \cite{bau14}), as well as for the sum of all the quantum contributions.

\begin{figure*}
    \centering
    \includegraphics[width=\textwidth]{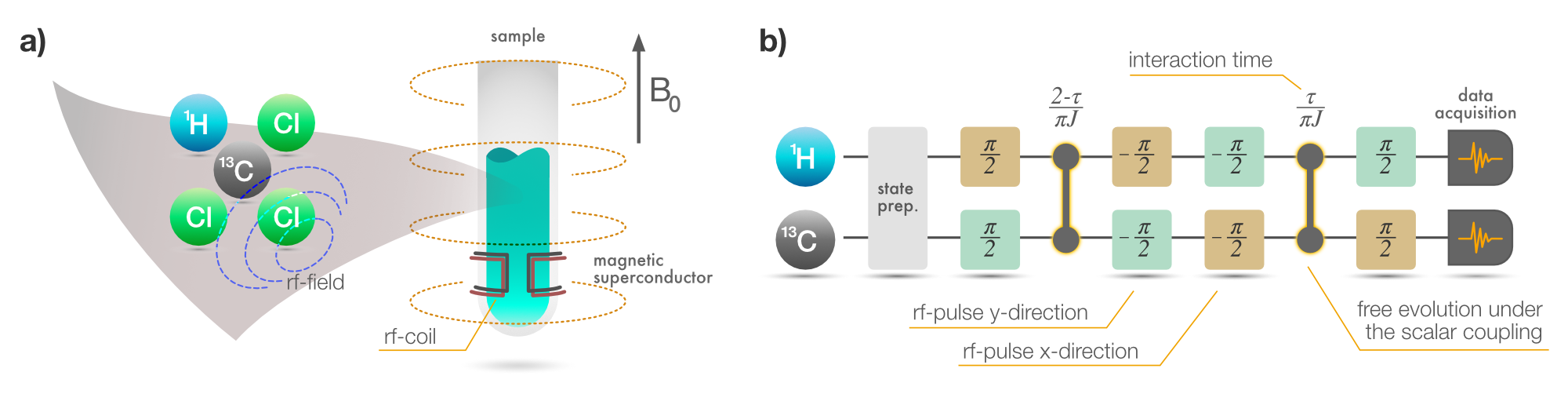}
    \caption{\label{fig:1}  Schematic  representation of the experimental system. a)  Two  qubits made of the nuclear spins-1/2 of  $^1$H  and $^{13}$C   of $^{13}$C-labelled chloroform  diluted in Acetone-d6 are placed in  a NMR magnetometer   that  produces a high intensity magnetic field
$(B_{0})$ in the longitudinal direction using
a superconducting magnet.  The sample is placed at the center of the magnet
within the radio frequency coil of the probe head inside a $5$mm
glass tube and immersed in a thermally
shielded vessel in liquid He, surrounded by liquid N in another vacuum
separated chamber. b) Experimental pulse sequence used to implement thermal interaction between the two qubits.
 The brown (green) square represents $x$ ($y$)
rotations by the indicated angle. The black vertical lines indicate
 free evolution under the scalar coupling, $\mathcal{H}_\text{J}^{\text{HC}}=({\pi\hbar}/{2})J\sigma_{z}^{\text{H}}\sigma_{z}^{\text{C}}$,
between the $^{1}$H and $^{13}$C nuclear spins with durations $2\tau/\pi J$ and $\tau/\pi J$. We  perform a total of 22 samplings of the interaction
time $\tau$ in the interval 0 to $2.32$~ms.
}
\end{figure*}
In our experiment, we consider two  qubits consisting of the nuclear spins-1/2 of  $^1$H (qubit $A$) and $^{13}$C (qubit $B$) from a $^{13}$C-labelled chloroform sample diluted in Acetone-d6. The sample is placed in a superconducting magnet that produces a static magnetic field in $z$-direction (Fig.~1a).  By combining transverse radio-frequency (rf) field with longitudinal field-gradient pulse sequences, we prepare an initial global state of the two spins-1/2 of the form (Methods),
\begin{equation}
\label{1}
    \rho^0_{AB} = \rho^0_A \otimes \rho^0_B + \chi_{AB},
\end{equation}
where $\chi_{AB} = \alpha\,|01\rangle\!\langle 10| + \alpha^*\,|10\rangle\!\langle 01|$ is a correlation term that satisfies $\mathrm{Tr}_j \chi_{AB} = 0$, $(j=A, B)$. As a result,  the initial local  states  are thermal, $\rho^0_j = \exp(-\beta_j H_j) / Z_j$,  with inverse  temperature $\beta_j$ and partition function $Z_j = \mathrm{Tr}_j \exp(-\beta_j H_j)$. This condition ensures
that the thermodynamic quantities of the local qubits are well-defined, even though they are globally correlated. The spin Hamiltonians are given in a double-rotating frame with the nuclear spins  Larmor frequency by $H_j = h\nu_0 (\mathbf{1} - \sigma_z) \big/2$, where $\sigma_z$ is the usual Pauli operator and $\nu_0 = 1\,\mathrm{kHz}$ is determined by the rf-field offset. We denote their  respective eigenstates by $|0\rangle$ and $|1\rangle$. To guarantee the positivity of the density operator $\rho^0_{AB}$, the correlation strength $\alpha$ should be  bounded by $|\alpha|\leq\exp[-h\nu_{0}(\beta_\text{A}+\beta_\text{B})/2]/({Z}_\text{A}{Z}_\text{B})$ \cite{mic19}. 
 The duration of the experiment (a few milliseconds) is much shorter than the decoherence time (a few seconds), so that the evolution of the global state can be considered as being unitary to an excellent degree of approximation \cite{bat14}.  The thermal interaction between the two qubits  is further realized via the exchange Hamiltonian $H_\text{int} = i ({\pi\hbar}/{2}) J \left( \sigma_A^+ \sigma_B^- - \sigma_A^- \sigma_B^+\right)$, where $J= 215.1\,$Hz. We implement the corresponding energy conserving evolution operator, $U_t = \exp({-i{t} H_\text{int}/{\hbar}})$ with $[U_t, H_A + H_B] = 0$,  by combining free evolution under the scalar coupling between $^1$H and $^{13}$C, and rf-field rotations (Fig.~1b).

\begin{figure*}
    \centering
    \includegraphics[width=1\textwidth]{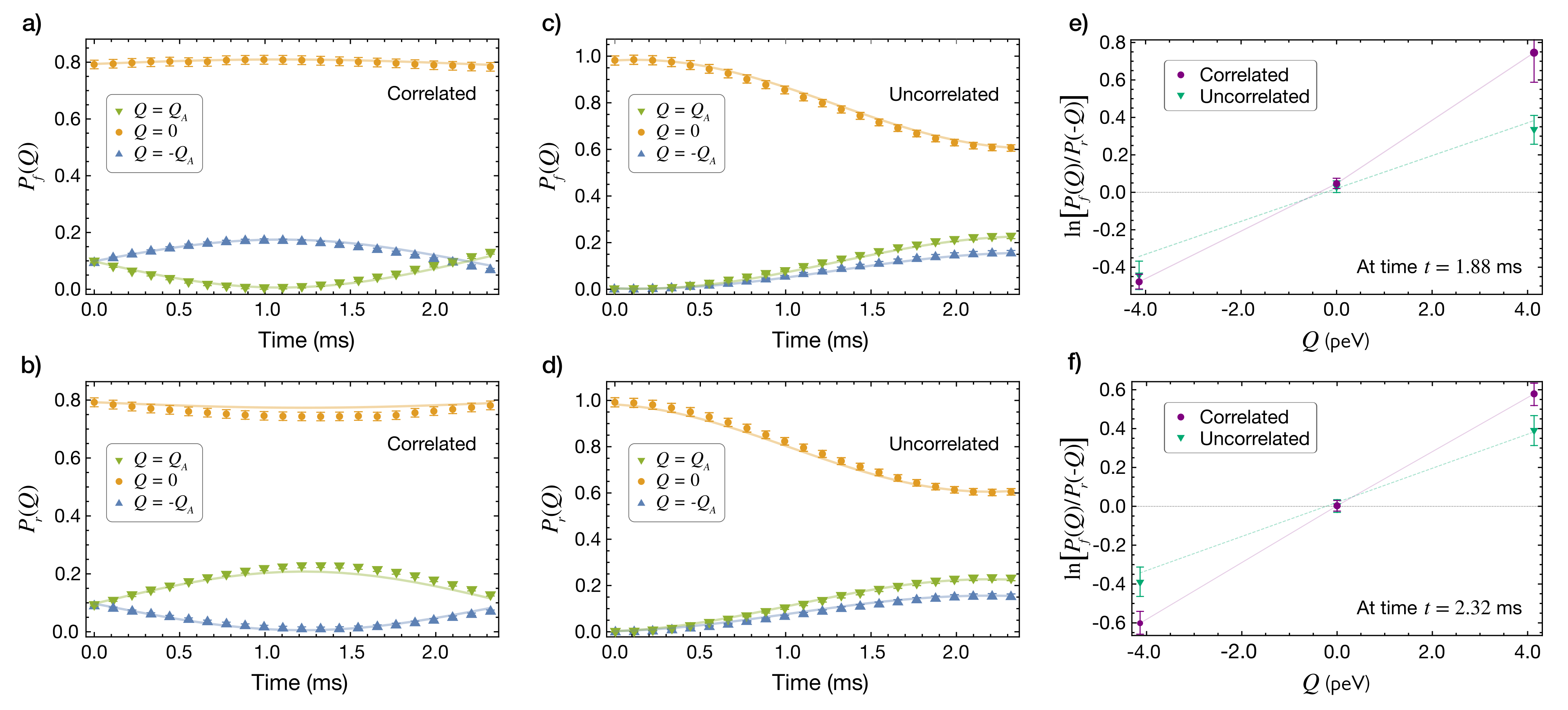}
    \caption{\label{fig:2} Detailed quantum fluctuation theorem with and without initial quantum correlations.
a) Forward and b) reverse heat distributions, $P_f(Q)$ and $P_r(-Q)$, as a function of time in the presence of initial quantum correlations between the two qubits. c) Forward and d) reverse heat distributions in the absence of initial correlations between the qubits.   Symbols represent data and  solid lines are  simulations \cite{mic20}. Error bars are evaluated by a Monte Carlo sampling of the standard deviation. e) Heat exchange fluctuation theorem, Eq.~(4), for $t= 1.88$ ms and f) for $t= 2.32$ ms, with (purple dots) and without (green triangles)  initial  correlations. In the absence of initial  correlations, the Jarzynski-W\'ojcik fluctuation relation ${P_f(Q)}/{P_r(-Q)} = \exp ( Q \Delta\beta)$  is obeyed, while the generalized theorem ${P_f(Q)}/{P_r(-Q)} = {\exp \left( Q \Delta\beta \right)}/{\Psi(Q)}$ that fully accounts for quantum correlations and quantum coherence with $\Psi(Q)\neq 1$ holds in their presence (solid and dashed lines are guide to the eye).}
\end{figure*}

Because of the nonzero correlations between the two qubits, the global state is not diagonal in the energy representation. Global and local bases are therefore  not mutually orthogonal, and the local bases, in which the exchanged heat variable is  evaluated, do not contain the entire information about the composite system. As a consequence, the two-projective-measurement scheme cannot account for quantum correlations and quantum coherence. A powerful approach that solves this incompatibility is provided by dynamic Bayesian networks \cite{nea03,dar09}. This formalism specifies the local dynamics conditioned on the global states, and hence preserves all the quantum properties of the system \cite{mic20}. By introducing conditional path trajectories for the two correlated systems and taking the average over the ensemble of all paths generated by the nonequilibrium heat exchange process leads to the integral quantum fluctuation theorem \cite{mic20},
\be
\label{2}
 \la \exp \left[-( Q_A \Delta\beta + I_0 - I_1 - \Sigma_A - \Sigma_B + \gamma  \right)]\ra = 1,
 \ee
 where $Q_A$ is the energy change of spin $A$ and $\Delta \beta = \beta_A - \beta_B$ the difference of inverse temperatures.
 In addition, $I_0$ $(I_1)$ is  the stochastic quantum mutual information that describes  initial (final) correlations between two subsystems  and $\Sigma_j$ is the stochastic relative entropy that  characterizes the entropy produced in spin $j$ (Methods). The last contribution $\gamma$ originates  from the random nature of the conditional dynamics, in analogy to the classical result of Ref.~\cite{sei05}. It vanishes on average,  since the global dynamics is unitary and no extra energy is exchanged with an external reservoir \cite{mic20}. Equation (2) shows that even in the absence of initial correlations, $I_0=0$, the two-projective-measurement scheme misses correlations, $I_1\neq0$, created during the heat exchange. 

Expression \eqref{2} generalizes the integral fluctuation theorem of  Jarzynski and  W\'ojcik, $\la \exp \left( Q_A \Delta\beta\right) \ra = 1$ \cite{jar04}. In order to highlight its quantum nature,  we write the stochastic quantum mutual informations, $I_l= J_l+ C_l$, $(l=0,1)$,  as a sum of the stochastic classical mutual information $J_l$  and of the stochastic quantum relative entropy of coherence $C_l$, which  is a proper measure of quantum coherence in a given basis \cite{bau14} (Methods).   The  fluctuation relation \eqref{2} thus fully captures the presence of quantum correlations between the two subsystems    and of quantum coherence in the global and local bases. Remarkably, contributions from both classical and quantum correlations, $J_l$ and $I_l$,   as well as from quantum coherence $C_l$, and the relative entropies $\Sigma_j$,  separately obey an integral fluctuation theorem \cite{mic20},
 \be
\label{3}
\langle e^{-I_l} \rangle=\langle e^{-J_l} \rangle=\langle e^{-C_l} \rangle=\langle e^{-\Sigma_j} \rangle= \langle e^{-\gamma} \rangle=1.
\ee
A detailed quantum fluctuation relation for heat may be similarly derived for the ratio of the forward heat distribution $P_f(Q)$ and its reverse distribution $P_r(Q)$ \cite{mic20},
\be
\label{4}
 \frac{P_f(Q)}{P_r(-Q)} = \frac{\exp \left( Q \Delta\beta \right)}{\Psi(Q)},
\ee
where the factor $\Psi(Q)$ depends on the initial correlations between the two qubits, such that the  Jarzynski-W\'ojcik result, $\Psi_\text{JW}(Q)=1$ is recovered in the absence of initial correlations between the two qubits \cite{jar04}.

In our experiment, in order to analyze the influence of correlations on the second law, we prepare the two-qubit system in an initial state of the form \eqref{1} with inverse spin temperatures $\beta_A^{-1} = 4.7(3)\,\mathrm{peV}$ $\big(\beta_A^{-1} =4.3(2)\,\mathrm{peV}\big)$ and $\beta_B^{-1} = 3.3(3)\,\mathrm{peV}$ $\big(\beta_B^{-1} = 3.7(3)\,\mathrm{peV}\big)$  with both $\alpha \neq0$ (initially correlated) and $\alpha=0$. We reconstruct the density matrix of the global state using  state tomography \cite{oli07} for a sequence of   22 values of time from $t=0$ to $t=\tau=2.32\,\mathrm{ms}$. We determine from these global states the respective local qubit states and all the relevant thermodynamic quantities appearing in the quantum fluctuation relations \eqref{2}-\eqref{4}. The thermal interaction $H_\text{int}$ induce four possible  transitions between the eigenstates of the two qubits.
This leads to three stochastic values of the heat, $Q = 0$ (twice) and $Q= \pm Q_A$, where $Q_A=  (E_{a_1}- E_{a_0})$ is the energy variation of spin $A$, with $E_{a_0}$ ($E_{a_1}$) the initial (final) energy eigenvalue of $H_A$. 

Figures 2a)-d) show the corresponding  forward heat distribution $P_f(Q)$ as well as its (time) reverse $P_r(Q)$ as a function of time, with ($\alpha= 0.17(1) + i 0.03(1)$) and without ($\alpha = -0.00(1) + i 0.0(1)$) initial correlations. We observe that the two heat distributions depend explicitly on time and that the forward and reverse distributions are identical in the absence of initial correlations. This follows from the fact that the global spin evolution is invariant under time reversal in that case \cite{jar04}. Figures 2e)-f) further exhibits the detailed heat fluctuation theorem \eqref{4} for two different times, $t= 1.88$ ms and $t = 2.32$ ms. Without initial correlations, we recover the Jarzynski-W\'ojcik relation in which corresponds to $ \ln [P_f(-Q)/{P_r(Q)}] = Q \Delta \beta$ is a straight line (green  triangles). For $\alpha \neq 0$, the effect of   quantum correlations  is clearly visible (purple  dots), modulating the $Q$-dependence via the function $\Psi(Q) \neq1$. Quantum correlations therefore modify  both the heat distributions and  the exponential dependence on the heat variable on the right-hand side of Eq.~\eqref{4} through the function $\Psi(Q)$. Two additional points  are noteworthy. First, since  correlations and coherence evolve with time during the heat exchange process, the function $\Psi(Q)$ is also time dependent, as seen in the two figures. Second, Fig.~2f) shows that the standard Jarzynski-W\'ojcik fluctuation theorem can be verified even with quantum correlations, albeit with a different slope. In this case,  quantum correlations and coherence lead to an effective inverse temperature difference which  is different from $\Delta \beta$.

The experimental study of the integral quantum fluctuation relations Eqs.~\eqref{2}-\eqref{3} is represented in Fig.~3. It reveals that not only the sum of all the contributions, $\sigma = -Q_A \Delta\beta -I_0 + I_1 + \Sigma_A + \Sigma_B - \gamma$,  in the exponent of Eq.~\eqref{2} satisfies a quantum fluctuation theorem, but that also individual contributions, $J_l$, $I_l$,   $C_l$,  $\Sigma_j$ ($l= 0, 1$ and $j=A, B$), and $\gamma$, separately obey such an integral  relation. These results are verified at all times and are illustrated for $t=1.77$ ms in the figure. Such findings suggest that many versions of the second law of thermodynamics hold independently, both 
for classical and quantum correlations, as well as for quantum coherence.

In summary, we have performed an extensive experimental investigation of fully quantum, detailed and integral, fluctuation relations based on a dynamic Bayesian approach. In contrast to the two-projective-measurement  method, such fluctuation theorems fully account for off-diagonal matrix elements in the local energy representation of a system, induced by either quantum correlations or quantum coherence. These improved formulations of the second law should therefore be useful for the study of the thermodynamic properties of small interacting quantum systems operating far from equilibrium.

\begin{figure}
    \centering
    \includegraphics[width=0.47\textwidth]{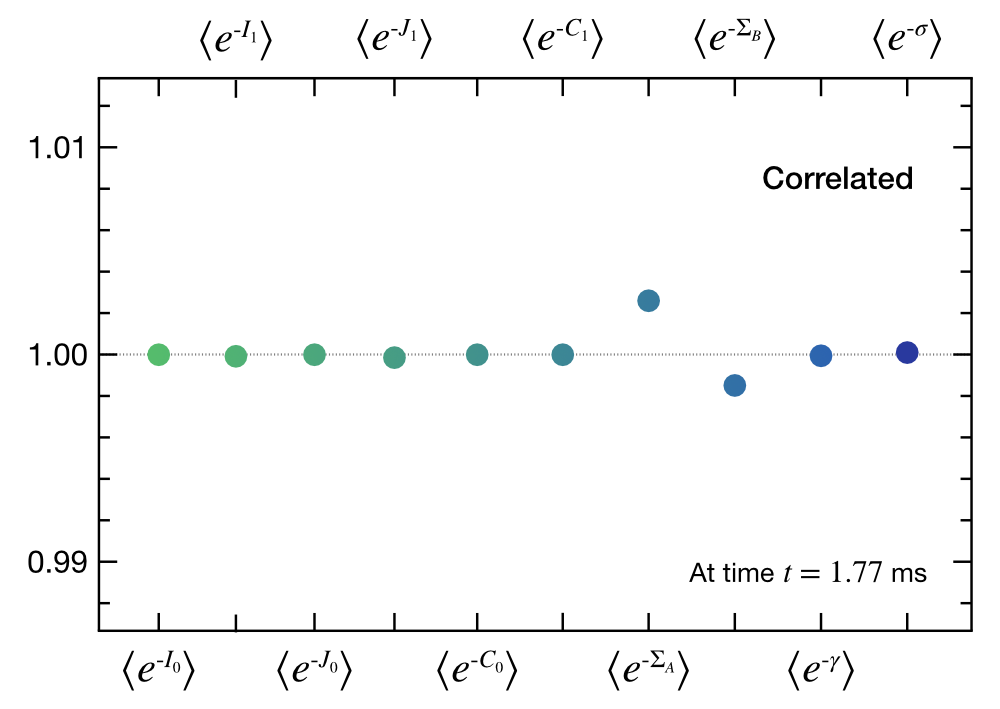}
    \caption{\label{fig:3} Integral fluctuation theorems with initial correlations.   The individual contributions from  classical and quantum correlations, $J_l$ and $I_l$,   as well as from quantum coherence $C_l$, and the relative entropies $\Sigma_j$,  for the two qubits ($l= 0, 1$ and $j=A, B$), and $\gamma$, separately verify the quantum integral fluctuation theorem (3). At the same time, the sum of all the quantum  contributions  $\sigma = -Q_A \Delta\beta -I_0 + I_1 + \Sigma_A + \Sigma_B - \gamma$ obeys the integral fluctuation relation (2).
}
\end{figure}

We acknowledge financial support from CNPq, CAPES, FAPERJ (Grant No. 203.166/2017), FAPESP (Grant No. 2016/08721-7 and 2018/12813-0),  the Ministery of Innovation, Science and Economic Development (Canada), the Government of Ontario, CIFAR, Mike and Ophelia Lazaridisthe, the S\~ao Paulo Research Foundation (Grants No. 2017/07973-5 and No. 2017/50304-7) and from the German Science Foundation (DFG) (Grant No. FOR 2724).  R.M.S.  gratefully acknowledges support from the Royal Society through the Newton Advanced Fellowship scheme (Grant No. NA140436) and the technical support from the Multiuser Experimental Facilities of UFABC. This research was performed as part of the Brazilian National Institute of Science and Technology for Quantum Information (INCT-IQ).
\section*{Methods}
\paragraph*{Experimental setup.} The experiment was performed in a Varian 500 MHz spectrometer equipped with a double-resonance probe head and a magnetic field-gradient coil. A 50 mg liquid sample of 99\% $^{13}$C-labeled CHCl$_3$ (Chloroform) was diluted in 0.7 ml of 99.9\% deutered Acetone-d6 and flame sealed in a 5 mm Wildmad LabGlass  tube.  Intermolecular interactions are negligible due to the high-level of dilution and  the system may considered as a set of identical pairs of spins-1/2. The superconducting magnet inside  the magnetometer produces a static longitudinal  magnetic field $B_0 \approx 11.75\,\mathrm{T}$, whose direction  is chosen as the positive $z$-axis. The respective Larmor frequencies of $^1$H and $ ^{13}$C are about 500 MHz and 125 MHz. 
The spin-lattice relaxation times, measured by the inversion recovery pulse sequence, are $\left(T_1^H, T_1^C\right) = (7.42, 11.31)$ s. Moreover, the transverse relaxations, obtained by the Carr–Purcell–Meiboom–Gill  pulse sequence, have characteristic times $\left(T_2^{*H}, T_2^{*C}\right) = (1.11, 0.30)$ s. 

\paragraph*{Thermodynamics.} Using the global and local decompositions, $\rho_{AB} = \sum_s P_s |s_n\rangle \langle s_n|$, $\rho_A = \sum_{a_n} P_{a_n}|a_n\rangle\langle a_n|$ and $\rho_B = \sum_{b_n} P_{b_n}|b_n\rangle\langle b_n|$, the probability of a  conditional path $\Gamma=(s,a_0,b_0,a_1,b_1)$  for a global unitary  $U_t$ is ${\cal P}(\Gamma) = P(s) \, |\langle a_0\,b_0 |s\rangle|^2 |\langle a_1\,b_1 | U_t |s\rangle|^2$.  The  forward heat distribution is equal to  $ P_f(Q) = \sum_{\Gamma} \delta(Q-Q_A)\, {\cal P}(\Gamma)$. Reverse distributions are similarly defined for the reversed path   $\Gamma^*=(s^*,a_1,b_1,a_0,b_0)$ \cite{mic20}.
We have additionally the two stochastic quantum mutual informations $I_0=\ln [{P_{s}}/{P_{a_0} P_{b_0}}]$ and $I_1=\ln [{P_{s^*}}/{\small \mathcal{P}(a_1) \mathcal{P}(b_1)}]$,  the two stochastic quantum relative entropies $\Sigma_A=\ln [{\small \mathcal{P}(a_1)/P_{a_1}}]$ and $\Sigma_B=\ln [{\small \mathcal{P}(b_1)/P_{b_1}}]$, as well as $\gamma = \ln [|\langle a_0\,b_0 |s\rangle|^2 |\langle a_1\,b_1 | U_t |s\rangle|^2/|\langle a_1\,b_1 | U^*_t |s^*\rangle|^2 |\langle a_0\,b_0 |s^*\rangle|^2)]$, with the probabilities  ${\small \mathcal{P}}(a_1) = \sum_{b_1} \langle a_1, b_1 | \rho_{AB}(t) | a_1,b_1\rangle$ and ${\small \mathcal{P}}(b_1) = \sum_{a_1} \langle a_1, b_1 | \rho_{AB}(t) | a_1,b_1\rangle$. The classical stochastic mutual information is further given by $J_l=\ln (P_{a_lb_l}/P_{a_l}P_{b_l})$, where $P_{a_lb_l}= \langle a_lb_l|\rho_{AB}^0|a_lb_l\rangle$, and  the stochastic quantum relative entropy of coherence reads $C_l= \ln (P_s/P_{a_l b_l})$.

 We determine the path probabilities $P(\Gamma)$ and $P(\Gamma^*)$ by diagonalizing the tomographically reconstructed global density matrices  to calculate the initial eigenvectors  $\{|s\rangle\}$, their probabilities $P_s$, as well as  the  evolved states $\{U_t|s\rangle\}$ and the local eigenstates from which we extract all the relevant thermodynamic quantities.

\clearpage

\section*{Supplementary Information}
\setcounter{figure}{0}
\renewcommand{\thefigure}{S\arabic{figure}}
\setcounter{table}{0}
\renewcommand{\thetable}{S\arabic{table}}
\setcounter{equation}{0}
\renewcommand{\theequation}{S\arabic{equation}}

\textit{Initial state preparation.} The initial states of the nuclear spins of $^1$H and $^{13}$C nuclei are prepared in local thermal states  by means of spatial average techniques \cite{Oliveira,Batalhao1,Batalhao2}. The sequence of pulses used to prepared the initial correlated state is depicted in Fig.~S1.\\

\vspace{5cm}

\begin{minipage}[c]{\textwidth}
\centering
 \hspace{-1.9cm}\includegraphics[width=6.4in]{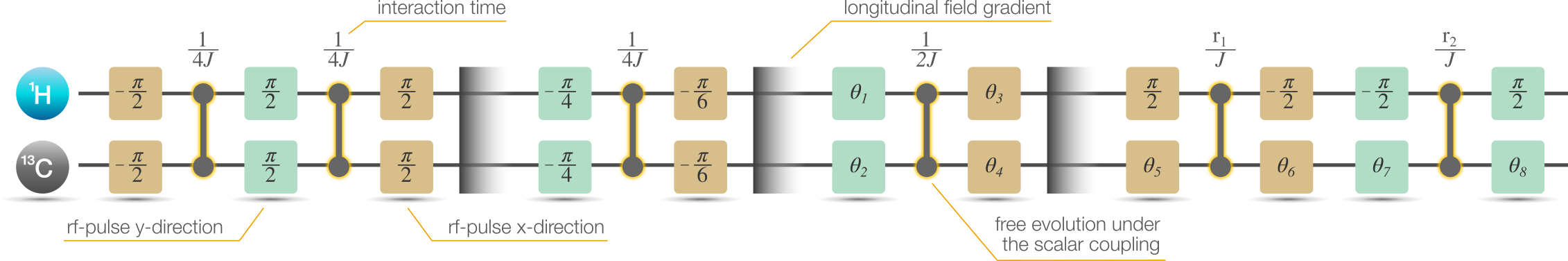}

\label{fig:sample_figure}
\end{minipage}
\begin{minipage}[t]{0.91\textwidth}
{Figure S1. The green (brown)  squares represent
local rotations by the indicated angle in $x(y)$-direction produced by a transverse rf-field resonant with $^1$H or the $^{13}$C nuclei adjusting phase, amplitude, and time duration. The vertical yellow connections represent the free evolution under the scalar coupling, $H^{HC} = ({\pi\hbar}/{2}) J\sigma_H\sigma_C$ ($J = 215.1$ Hz), between the $^1$H and $^{13}$C nuclear spins along the time indicated above the symbol. In order to build the initial state equivalent to the one described in Eq.~(1) of the main text, the modulation and intensity of the gradient pulse are optimized, as well as the angles $\{\theta_1,\dots , \theta_8\}$ and the interaction times $r_1$ and $r_2$.}\end{minipage}

\end{document}